\documentclass[11pt,twoside]{article} 
\usepackage{asp2004}
\usepackage{epsf}
\usepackage{psfig}
\usepackage{lscape} 

\begin{document} 
\title{White Dwarfs in $\omega$ Centauri: Preliminary Evidence}

\author{G. Bono,$^1$ M. Monelli,$^{1,2}$ P. B. Stetson,$^3$ P. G. Prada
Moroni,$^4$ R. Buonanno,$^{1,2}$ A. Calamida,$^{1,2}$ I. Ferraro,$^1$
G. Iannicola,$^1$ C. E. Corsi,$^1$ and L. Pulone$^1$}

\affil{$^1$INAF-Osservatorio Astronomico di Roma, via Frascati 33,
00040 Monte Porzio Catone, Rome, Italy\\ 
$^2$Universit\`a di Roma Tor Vergata, via della Ricerca Scientifica 1,
00133 Rome, Italy\\ 
$^3$Dominion Astrophysical Observatory, 5071 West Saanich Road,
Victoria, BC V9E 2E7, Canada\\ 
$^4$Dipartimento di Fisica "E. Fermi", Univ. Pisa, via B. Pontecorvo
2, 56127 Pisa, Italy\\}


\begin{abstract} 
We present accurate and deep multiband B,R,$H_\alpha$ data for the globular 
cluster \hbox{$\omega$~Cen} collected with the Advanced Camera for Surveys on 
HST. The photometric catalogue includes more than one million stars. 
By adopting severe selection criteria we identified more than 600 bona fide 
White Dwarfs (WDs). Empirical evidence suggests that a small sample of WDs 
are H$_\alpha$-bright. The comparison between WD isochrones and observations 
shows a reasonable agreement at fainter magnitudes and a mismatch at the 
brighter ones.     
\end{abstract}

\section{Introduction}
The Galactic Globular Cluster (GGC) \hbox{$\omega$ Cen~} is a
fundamental laboratory to address several long-standing astrophysical
problems. It is the most massive GGC ($M=5\times10^6\, M_\odot$,
Meylan et al.\ 1995) and the one that most clearly shows a
well-defined spread in metallicity. According to recent estimates
based on sizable samples of evolved red giant and sub-giant stars, the
metallicity distribution shows three peaks around ${[\rm Fe/H}]=-1.7$,
$-1.5$, and $-1.2$ together with a tail of metal-rich stars
approaching ${[\rm Fe/H}]\approx-0.5$ (Norris et al.\ 1996; Hilker et
al.\ 2004; Pancino 2004). During the last few years it has also been
suggested that \hbox{$\omega$ Cen~} harbors multiple stellar
populations (Lee et al.\ 1999) characterized by different ages
(Ferraro et al.\ 2004; Hughes et al.\ 2004), helium abundances (Bedin
et al.\ 2004; Norris 2004), and distances (Bedin et al.\ 2004;
Freyhammer et al.\ 2004). From a kinematical point of view the
properties of \hbox{$\omega$ Cen~} appear well-established: it moves
along a high-eccentricity, retrograde orbit (Geyer et al.\ 1983), and
shows differential rotation (Merritt et al.\ 1997). The occurrence of
a tidal tail in \hbox{$\omega$ Cen~} was suggested by Leon et al.\
(2000) but questioned by Law et al.\ (2003) on the basis of 2MASS
data. This problem has not been settled yet, because recent detailed
N-body simulations of the tidal interaction between \hbox{$\omega$
Cen~} and the Galaxy do suggest the occurrence of extended tails
(Chiba \& Mizutani 2004).  Current empirical and theoretical evidence
do not allow us to establish whether \hbox{$\omega$ Cen~} is the core
of a galaxy that was partially distrupted by the gravitational
interaction with the Galaxy (Lee et al.\ 1999; Pancino 2004) or the
aftermath of the merging of two GCs (Icke \& Alcaino 1988).

Even though \hbox{$\omega$ Cen~} presents several properties that need to be properly 
understood, its stellar content is a gold mine to investigate some open 
problems concerning the dependence on the metallicity. This applies not 
only to evolved stars such as RR Lyrae, hot HB stars, and the tip of 
the RG branch, but also to the different expected population(s) of 
white dwarfs. The search for WDs in GGCs has been quite successful, 
and several cooling sequences have already been identified 
(Hansen et al.\ 2002; Moehler et al.\ 2004, and references therein).   
The detection of WDs in \hbox{$\omega$ Cen~} dates back to  Ortolani \& 
Rosino (1987) and to Elson et al.\ (1995) on the basis of 
ground-based and space (HST) data, respectively. In this paper, 
we present preliminary results concerning the identification 
of the WD cooling sequence in \hbox{$\omega$ Cen~} on the basis of  
data collected with the Advanced Camera for Survey (ACS) on board 
HST, and publicly available on the HST archive.

\section{Data Reduction and Color-Magnitude Diagrams}

Current data were collected with nine pointings of the ACS camera
across the center of the cluster. The $3\times3$ mosaic covers a field
of view of $\approx9\arcmin \times9\arcmin$.  Four images per field
have been acquired in three different bands, namely F435W, F625W, and
F658N (hereinafter $B$, $R$, and $H_\alpha$).  Three deep (340s) and
one shallow (8s, 12s) exposure were secured for the B and R-band,
respectively, while the exposure time for the four $H_\alpha$ images
was 440s. The nine fields were independently reduced with the
DAOPHOTII/ALLFRAME package (Stetson 1994). An individual PSF has been
extracted for each frame by adopting, on average, $\approx$200 bright
isolated stars.  The individual catalogues were rescaled to a common
geometrical system with DAOMATCH/DAOMASTER. The final catalogue
includes approximately 1.2$\times 10^6$ stars. The photometric
calibration was performed in the Vega System
(http://www.stsci.edu/hst/acs/documents). The same data set was
adopted by Haggard et al.\ (2004) to identify the optical counterpart
to a quiescent neutron star originally detected on X-ray data
collected with Chandra.

\begin{figure}[!ht] 
\vspace*{0.5truecm}
\centerline{\epsfxsize= 9.0 cm \epsfysize= 10.0 cm \epsfbox{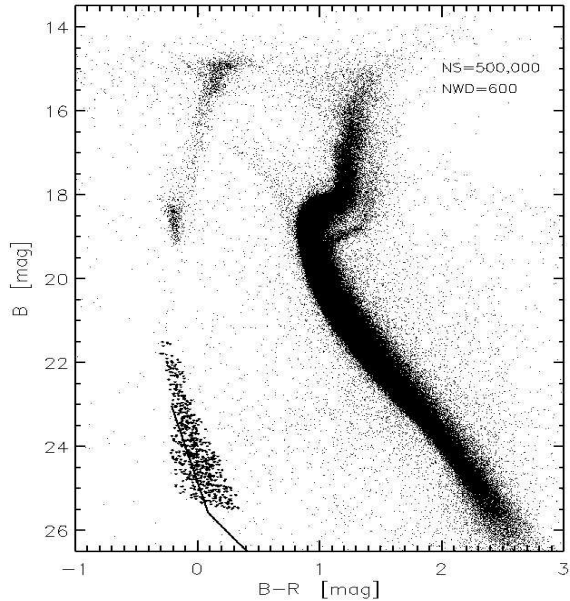}}
\vspace*{-0.3truecm}
\caption{Color-Magnitude Diagram, $B,B-R$, based on data collected
with ACS@HST.  The photometric catalogue was selected by assuming a
separation index $sep \ge 0.1$.  Thick points display WDs with
photometric errors smaller than 0.2 mag. The solid line shows a WD
isochrone for t=14 Gyr and Z=0.0001. See text for more details.}
\end{figure}

Figure 1 shows the CMD of \hbox{$\omega$ Cen~} in the $B-R, B$ plane. Data plotted in
this figure have been selected by using the `separation index' {\em
sep} introduced by Stetson et al.\ (2003). We adopted this parameter
($sep \ge 0.1$) because crowding errors in the innermost regions of
GGCs dominate the photometric errors.  The WDs candidates have been
selected among the objects with photometric errors in the three bands
smaller than 0.2 mag. We end up with a sample of half a million cluster
stars and roughly 600 WDs (thick points). To compare observed WDs with
theoretical predictions we adopted the recent WD theoretical models
computed by Prada Moroni \& Straniero (2002). Predicted luminosities
and effective temperatures have been transformed into the ACS bands by
adopting the pure H atmosphere models kindly provided by Bergeron
(private communication). The solid line plotted in Fig. 1 shows the WD
isochrone for t=14 Gyr and Z=0.0001.  We adopted the same distance
modulus (DM=13.7) and cluster reddening (E(B-V)=0.12) adopted by
Freyhammer et al.\ (2004). The extinction in the ACS bands was
estimated using the Cardelli et al.\ (1989) relation. Theory and
observations appear to agree at fainter magnitudes and cooler
effective temperatures. However, we are faced with a substantial
discrepancy in the bright region. At present it is not clear whether
this mismatch is caused either by the assumed composition of the
atmosphere models, or by the inner structure of the brightest WD
models.  More quantitative constraints do require a thorough
comparison between theory and observations.

To further investigate the evolutionary properties of cluster WDs we plotted 
the same sample in the $H_\alpha-R, R$ plane (see Fig. 2). 

\begin{figure}[!ht] 
\vspace*{0.5truecm}
\centerline{\epsfxsize= 9.0 cm \epsfysize= 10.0 cm \epsfbox{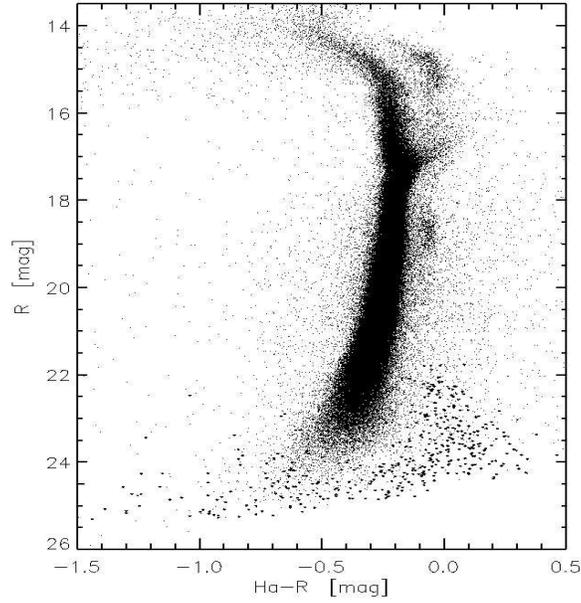}}
\vspace*{-0.3truecm}
\caption{Same as Fig. 1, but for $H_\alpha$ and R-band photometry.}     
\end{figure}

This plane is generally adopted to detect stars with strong $H_\alpha$
emission, namely Cataclysmic Variables (CVs) or BY Draconis stars.
The former group is characterized by flicker variations, while
the latter group contains chromospherically active Main Sequence
stars (dK,dM) with variability of the order of days caused by
fast rotation. Recent photometric (Cool et al.\ 1998) and
spectroscopic (Edmonds et al.\ 1999) measurements of a few blue stars
in the GGC NGC6397 suggest that He WDs also show $H_\alpha$ emission,
but they lack the flickering variations. Needless to say, that He WDs
are excellent tracers of the dynamical properties of GGCs, since they
are the aftermath of compact binary evolution (Taylor et al.\ 2001).
Data plotted in Fig. 2 show that the WD sequence detected in the
$B-R, B$ plane shows up also in this plane. The broadening in color when
moving from $R\approx 22$ to $R\approx 24.5$ is mainly caused by
photometric errors.

\begin{figure}[!ht]  
\centerline{\epsfxsize= 10.5 cm \epsfbox{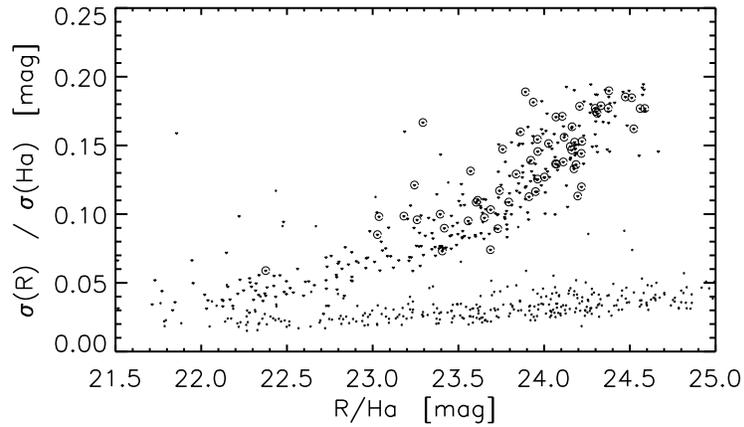}} 
\vspace*{-0.3truecm}  
\caption{Intrinsic photometric error in $R$ (thin points) and $H_{\alpha}$ 
(thick points) bands. Open circles mark WDs with $H_{\alpha}-R\le -0.2$, and 
$|sharp|\le0.1$.}     
\end{figure}

However, data plotted in this plane show quite clearly that a small sample  
of WDs are $H_\alpha$-bright. The identification of these objects 
is quite easy, since they show R magnitudes of the order of $R\approx 24-25$ 
but they appear systematically bluer than typical WDs. It is noteworthy 
that the identification of $H_\alpha$-bright WDs becomes 
more robust when moving toward bluer colors. In order to supply a more 
quantitative estimate of the photometric errors affecting current mean 
magnitudes, Fig. 3 shows the standard deviations for the R (thin points) 
and the $H_\alpha$ (thick points) mean magnitudes. To further improve the photometric 
characterization of $H_\alpha$-bright WDs we selected (open circles) the WDs 
with $H_\alpha-R \le -0.2$ and $|sharpness|\le 0.1$ (this parameter estimates 
the sharpness of the detected object). We end up with a sample of $\sim 70$ 
candidates. Data plotted in Fig. 3 show that typical errors for these objects 
range from less than 0.1 to $\sim 0.2$ mag when approaching the faintest 
limiting magnitude.

\begin{figure}[!ht] 
\vspace*{0.5truecm}
\centerline{\epsfxsize= 9.0 cm \epsfysize= 10.0 cm \epsfbox{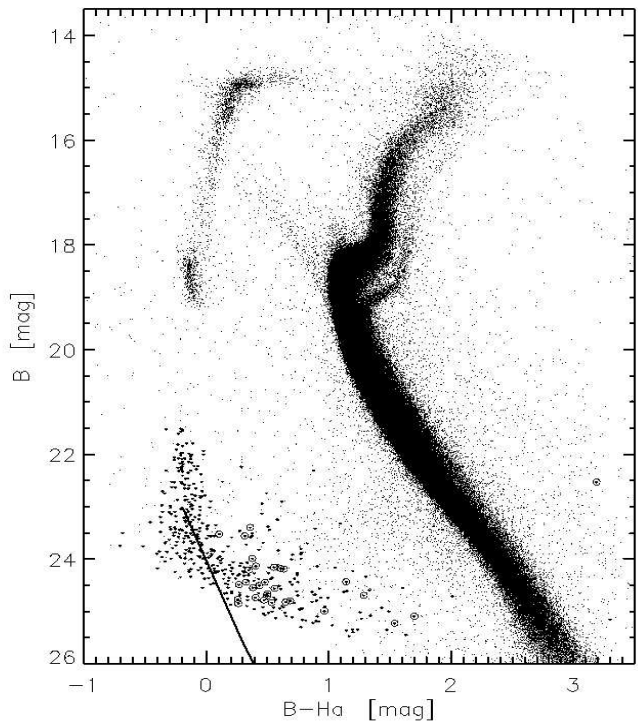}}
\vspace*{-0.3truecm}
\caption{Same as Fig. 1, but for $B,B-H_{\alpha}$ photometry.     
Open circles mark WDs with $R-H_{\alpha}\ge 0.2$, and $|sharpness|\le0.1$.}     
\end{figure}

Finally, we plotted the entire sample of WDs in the $B-H_{\alpha}, B$ plane. 
Once again we found that a small sample appears to be 
systematically brighter in $H_{\alpha}$ (open circles), since they are  
redder than the bulk of the WDs and of the WD isochrone (solid line).     


Current data indicate that a small sample of bonafide WDs are 
$H_{\alpha}$-bright. These objects do not appear to be located at 
a fixed color range in the $B-R, B$  plane. This evidence suggests that 
this phenomenon is not correlated with the occurrence of a circumstellar 
envelope caused by pulsation properties of DA and DB pulsating WDs 
(Nitta et al.\ this conference).  

Moreover, if independent photometric or spectroscopic measurements
confirm that $H_{\alpha}$-bright WDs are truly He WDs, the size of the
sample would pose a new puzzle concerning their origin. According to
recent N-body simulations the occurrence of compact binaries is
tightly connected with the dynamical evolution of the cluster and they
should peak in post-core collapsed clusters, such as NGC6397. However,
\hbox{$\omega$ Cen~} presents a low central density.

\acknowledgements{This work was partially supported by MIUR-COFIN~2003 under 
the project "Continuity and Discontinuity in the Galaxy Formation".}

\end{document}